# Metallicity and conductivity crossover in white light illuminated $CH_3NH_3PbI_3$ perovskite

A. Pisoni[1], J. Jacimovic[1], B. Náfrádi[1], P. Szirmai[1], M. Spina[1], R. Gaál[1], K. Holczer[2], E. Tutis[3], L. Forró[1*] and E. Horváth[1]

[1]Laboratory of Nanostructures and Novel Electronic Materials, EPFL, Lausanne, CH-1015, Switzerland
[2]Physics Department, UCLA, Los Angeles, CA 90995-1547, USA
[3]Institute of Physics, Zagreb, Croatia

**The intrinsic d.c. electrical resistivity ($\rho$) - measurable on single crystals only – is often the quantity first revealing the properties of a given material. In the case of $CH_3NH_3PbI_3$ perovskite measuring $\rho$ under white light illumination provides insight into the coexistence of extended and shallow localized states (0.1 eV below the conduction band). The former ones dominate the electrical conduction while the latter, coming from neutral defects, serve as a long-lifetime charge carrier reservoir accessible for charge transport by thermal excitation. Remarkably, in the best crystals the electrical resistivity shows a metallic behaviour under illumination up to room temperature, giving a new dimension to the material in basic physical studies.**

$CH_3NH_3PbI_3$ belongs to the family of methylammonium-trihalogeno-plumbates, discovered by Weber in 1978[1]. Recently, a strong interest in this material was triggered by the reports of the Snaith [2,3] and Graetzel [4,5] groups of rather simple photovoltaic devices reaching a light conversion efficiency of 15-16%, using relatively undemanding synthesis conditions. Further on, the material was found to be promising for a bright and cheap light emitting diode[6], it has shown lasing at relatively low pumping threshold[7], and opened the possibility for the first sustainable water splitter[8].

The great performance of $CH_3NH_3PbI_3$ in photovoltaics (PV) is still an enigma. The high efficiency of a PV cell needs good light absorption capacity, long diffusion length of the created excitons, and good mobility of the emerging electrons and holes. These conditions being satisfied, the high quantum efficiency (electron per photon) and power efficiency (device voltage vs. energy of absorbed photon) emerge almost independently of device architecture. The two last conditions require a material with low scattering, trapping, and (in-trap) recombination rates. Recent measurements have indeed uncovered that the electrons and holes (combined into excitons and individually) propagate from their creation spot to distances of the order of 0.1-1 μm [3]. Nevertheless, the high performance of $CH_3NH_3PbI_3$ is very surprising for a material prepared from solution through simple spin-coating process. Finding the origin of the stunning performance of this cheaply prepared crystal is not just a matter of single material or application. It is a physical puzzle of its own, with possible wider implications on future directions in materials design.

The understanding of the charge transport of photo-induced carriers is an important step in the description of the material. The resistivity measurements on single crystals reported below demonstrate the interplay of extended and localized states in the material, determine the energy scale which separates them, and reveals the hierarchy of processes relevant for photoconduction.

All measurements were performed on $CH_3NH_3PbI_3$ (hereafter $MAPbI_3$) single crystals. The samples (see Fig. 1 and Methods) were mounted on a cold finger of a cryo-cooler configured with an optical window and the resistance was measured as a function of temperature under constant white light illumination provided by the light source of an optical microscope lamp (see Fig. 1b). The nominal intensities were up to $59*10^3$ lux. (Note that the average outdoor sunlight is in the $32*10^3 - 100*10^3$ lux range).

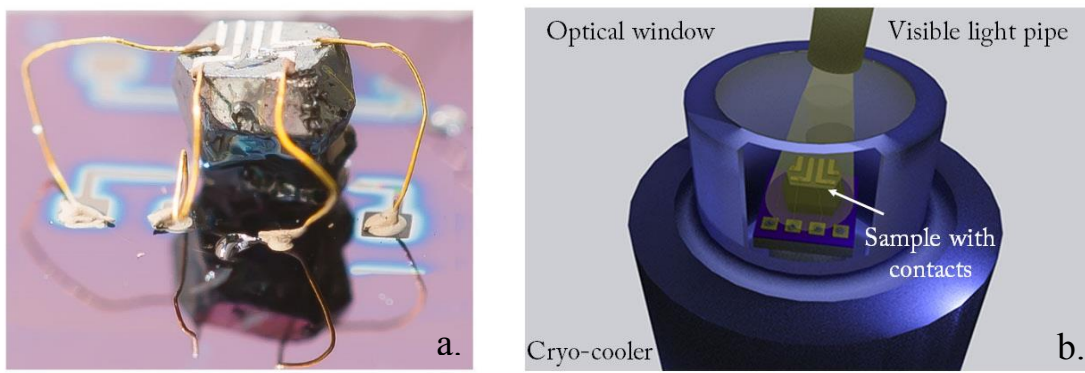

**Figure 1 | Configuration for resistivity measurement of MAPbI$_3$ single crystal under white light illumination. a,** The single crystal placed on a SiO$_2$ support is contacted by gold wires for 4 point resistivity measurement. The longitudinal dimension of the sample is in the range of 5 mm. **b**, Sketch of the experimental configuration of the temperature dependent resistivity measurement under illumination. The white light is introduced via an optical window of a cryo-cooler.

Figure 2 shows the I-V characteristics of a single crystal of MAPbI$_3$ at room temperature with increasing white light intensity, which was chosen to be closer to the working conditions of photovoltaic devices. (The effect was qualitatively identical for monochromatic red laser light). The I-V curves are symmetric for positive and negative biases. The photocurrent at fixed bias (1 V) rapidly increases at low light intensities as shown in the inset of Figure 2 and doesn't saturate up to the highest intensity available. The very high photocurrent is one of the key reasons for the high efficiency of the photovoltaic cells constructed with MAPbI$_3$. The resistivity response to white light illumination is very strong: from dark to the maximum light intensity (59*10$^3$ lux) it drops by a factor of 100 to 2000, depending on the crystal quality.

This strong response to illumination is magnified with decreasing temperatures. In Fig. 3a, the temperature dependence of the resistivity of a single crystal (average quality, handled without special care) is shown for 0, 16, 33 and 44 thousand lux (denoted as dark, min, mid and max intensities, respectively). The Arrhenius plot of data in the high temperature region, shown in Fig. 3b, demonstrates the temperature-activated behaviour and reveals the energy scales involved. The activation energies are: 185±10 meV for the sample in dark and in the 120±10 meV range for illuminated conditions. At this point one has to note the extraordinary high value of the dark resistivity[9]. The room temperature (extrapolated) value is about $2 \times 10^7$ $\Omega$cm: this is about 2 orders of magnitude higher than the intrinsic resistivity for Silicon, and 5 orders of magnitude higher than Phosphorus doped Silicon at $10^{13}$ cm$^{-3}$ doping concentration[10]. While the small activation energy clearly indicates the presence of extrinsic carriers, there are very few of them: the "as grown crystals" appear to be extremely pure! We estimate an upper bound for the extrinsic carrier density to be less than $10^{13}$ cm$^{-3}$ [11]. Such "purity levels" have been difficult to achieve with elemental semiconductors, where inclusion of almost any other atom leads to a charged defect. It is very unlikely that MAPbI$_3$ is so perfect but rather that the defects are special. The complex nature of MAPbI$_3$ allows and favours the formation of a variety of charge-neutral defects preserving strongly local charge neutrality – a distinguishing feature from classical semiconductors.

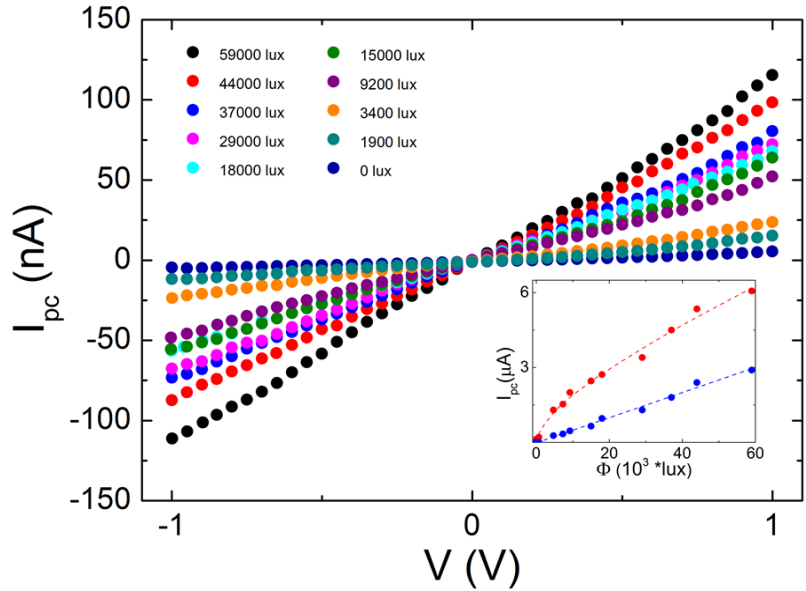

**Figure 2 | Photo-induced current ($I_{pc}$)-voltage characteristics of MAPbI$_3$ single crystal as a function of white light intensity ($\Phi$) measured at room temperature.** The inset shows the photo-current variation with $\Phi$ at fixed bias voltage of 1 V. The continuous line is a fit to the microscopic model in Eq. 4. Note that the linear plus square root dependence of $I_{pc}$ on $\Phi$ at 300 K is quite unique for a photovoltaic material. For the sake of comparison, it is given at 50 K, as well where it shows linear dependence on $\Phi$. Both regimes are fully reproduced by the model developed below.

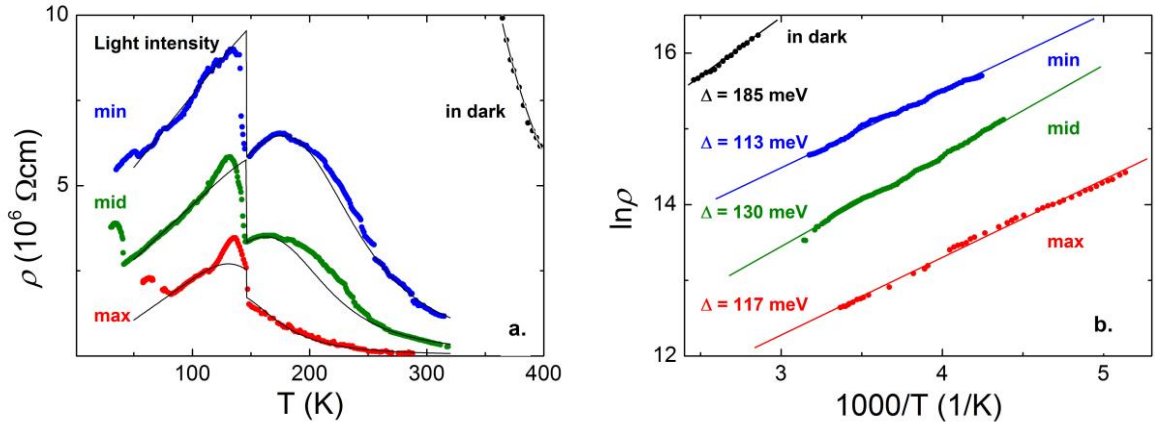

**Figure 3 | Temperature dependence of the resistivity of a MAPbI$_3$ single crystal. a**, in dark and under illumination with three white light intensities ranging from the lowest (min: 16*10$^3$ lux) through the intermediate (mid: 33*10$^3$ lux) to the highest (max: 44*10$^3$ lux) ones; The jump in the resistivity at 150 K is due to the structural phase transition; (the continuous lines are coming from the fit developed in the model below; the fit parameters are given in the supplementary information) **b**, Arrhenius plot of the high temperature semiconducting part with extracted activation energies is shown.

The activation energy is markedly lower for the illuminated samples than for the dark one, pointing to the possibility of different type of states being involved. The activated temperature dependence suggests the possibility of photo-induced "dopant states" or the presence of dynamically bound, shallow states of photo-induced carriers (e.g. excitons). In both cases the effective number of carriers in the band, contributing to electric conduction, diminishes as the temperature decreases. The experiment, however, tells us that the situation is significantly more interesting.

Upon cooling below a certain temperature $T_x$ the resistivity starts to decrease. This crossover from "non-metallic" to "metallic" temperature dependence of the resistivity is the general property of our samples, although the crossover temperature $T_x$ depends on the illumination intensity (Fig. 3a) and on the age of the sample. It is known that with time the sample quality is diminished, as shown in Fig. 4 for crystals of different age (from freshly prepared #3 to several days old #1). It is evident that $T_x$ varies greatly between batches, while the temperature of the structural phase transition remains almost unchanged. $T_x$ is unrelated to the structural phase transition[12,13] which is always observed around $T_c$ =150 - 160 K (i.e. does show weak ageing dependence, see Fig 4). The structural transition shows up as a pronounced jump in resistivity at $T_c$, (Fig. 3a), but does not affect the overall behaviour of the resistivity vs. temperature. The small jumps in ρ in certain runs are believed to be due to stress induced micro-cracks resulting from the 2.5 % decrease in the unit cell volume at the structural transition[14,15].

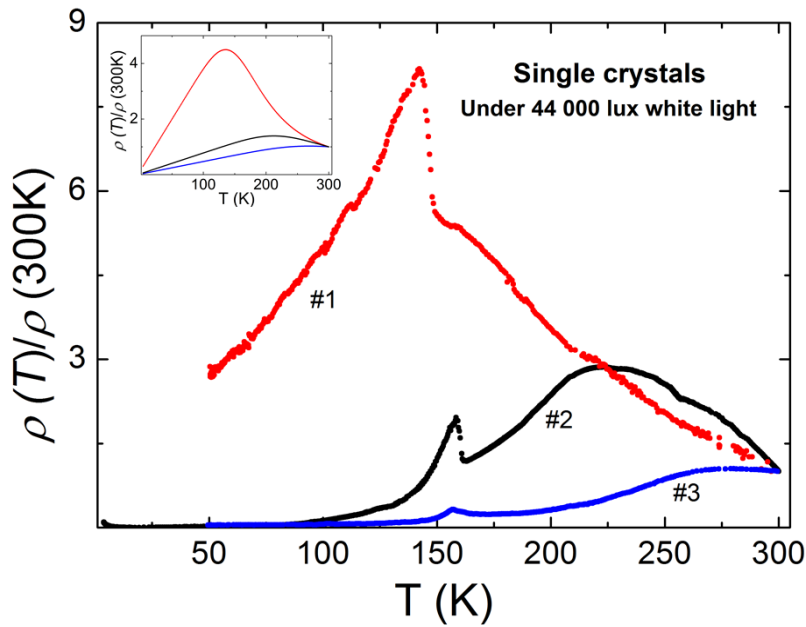

**Figure 4 | Sample dependence of the resistivity of MAPbI$_3$ single crystals under white light illumination.** The three samples were illuminated with 44*10$^3$ lux and the resistances were normalized to their room temperature value. All samples show striking metallic behaviour over some temperature range, but the range of the semiconducting behaviour varies considerably. It is attributed to the density of localized states which evolves by aging of the samples (type-2 in the text). The inset represents the expected variation of ρ for c=c$_1$ (blue), c=3 c$_1$ (black), c=20 c$_1$ (red) in Eq. (4). With increasing c the spread of localized states increases and as a consequence the activation energy Δ decreases. In the simulations of the inset it has the following values: blue:130 meV, black: 120 meV, red:100 meV. The room temperature values of the resistivities (calculated for a thickness of light absorption of 1 μm,) are: blue:5.7*10$^1$ Ωcm, black: 4.2*10$^3$ Ωcm, red: 5.0*10$^4$ Ωcm.

The crossover from 'semiconducting' to 'metallic' behaviour is highly unexpected. The "metallic" resistivity in illuminated samples is a clear sign of charge carriers moving in extended band states, with the temperature dependence of scattering similar to ordinary metals. The apparent disappearance of the effect of "localized" or bound states in the low temperature region may be understood as the sign of them being full/saturated in the low temperature range. The saturation is not expected for dynamically bound states, nor is their sample/age dependence. The picture of "photo-induced dopants" has similar problems. In fact, it is natural to assume that (white light) illumination places the charge carriers mostly into the band of extended states, rather than into localized "dopant" states. This shifts the question of carrier density towards carrier dynamics rather than equilibrium statistics. Bellow we present the simple model that addresses the questions raised by the experiment[16].

Naturally, the model builds upon the interplay of two types of electronic states: extended band states (which we refer to as type-1 states) that dominantly contribute to electrical conductivity and localised (type-2) states at energy $-\Delta$ with respect to the bottom of the conduction band (CB) [16]. The latter states do not contribute directly to the photo-conductivity, but play the role of the charge reservoir that communicates with band states. The model is pictured in Fig. 5. Upon illumination, the states are populated by photo-excited carriers. These carriers are excited from the valence band (VB) well above the charge transfer gap ($\Delta_0 = 1.5$ eV)[17], further relaxing through transitions to extended and localized states. Unless they reach an electrode, as in a photovoltaic device, these carriers ultimately recombine (radiatively or non-radiatively), possibly going through intermediate (singlet or triplet) excitons. Our photo-conductivity experiments do not speak much about excitons (being charge neutral, they are "mute" in charge transport), but give good insight into the steady-state population of carriers in extended states and their interplay with localized/mute states.

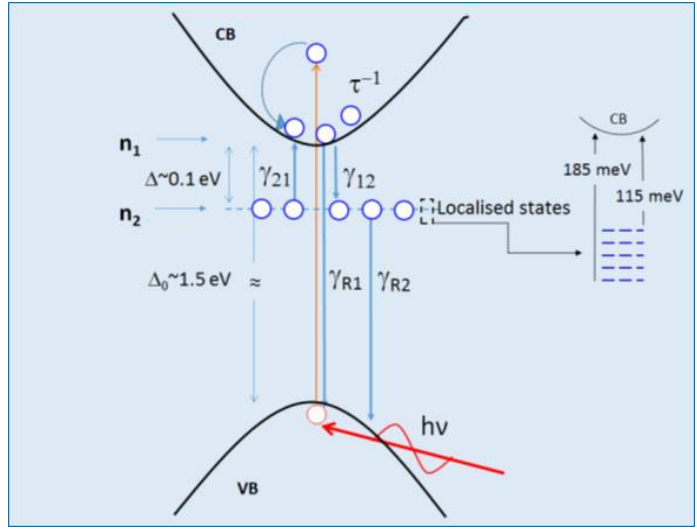

**Figure 5 | Sketch of the electronic states of MAPbI$_3$.** The major quantities used in the microscopic modelling of MAPbI$_3$ are the following: $n_1$ and $n_2$ are the occupancies of type-1 (extended) and type-2 (localised) states. The transfer rates between these states are: $\gamma_{12}$ and $\gamma_{21}$, and the recombination rates from them are $\gamma_{R1}$ and $\gamma_{R2}$. The scattering rate in type-1 state is $\tau_1^{-1}$. The blue and red circles denote electrons and holes, respectively created by the photon impact. CB and VB stand for conduction and valence band, respectively. The blow-up of the localised states shows the spread in energy of ~70 meV based on Fig. 3b: they are populated by photo-excited electrons and the activation energy changes relative to the dark case.

The model includes the generation of carriers by light, the recombination that depopulates both type of states, and transitions between the type-1 and type-2 states. The rate equations for the occupancy of the first and second type of states, denoted by $n_1$ and $n_2$ respectively, are

$$\frac{dn_1}{dt} = \Phi - \gamma_{R1} n_1^2 - c\gamma_{12} n_1 + c\gamma_{21} n_2 , \qquad (1)$$

$$\frac{dn_2}{dt} = -\gamma_{R2} n_2^2 + \gamma_{12} n_1 - \gamma_{21} n_2 . \qquad (2)$$

For simplicity, we assume that carriers are excited only into extended band states (a more detailed model is presented in the supplement). The excitation rate, being proportional to the light intensity $\Phi$. The concentration of type-2 states is given by $c$. The terms $\gamma_{R1} n_1^2$ and $\gamma_{R2} n_2^2$ stand for the recombination of electrons and holes ($p$) in type-1 and type-2 states, respectively (with concentrations of electrons $n_{1,2}$ and holes $p_{1,2} \approx n_{1,2}$, and the pair-forming probabilities given by $\gamma_{R1}n_1p_1 \approx \gamma_{R1}n_1^2$ and $\gamma_{R2}n_2p_2 \approx \gamma_{R2}n_2^2$)[18]. The rate constants $\gamma_{12}$ and $\gamma_{21}$ are used to model the transitions between the two types of states. The energy difference Δ between type-1 and type-2 states enters into the microbalance requirement relating $\gamma_{21}$ and $\gamma_{12}$,

$$\gamma_{21} = \gamma_{12}\, e^{-\frac{\Delta}{T}}. \tag{3}$$

This is the only mandatory temperature dependence in the model, and the only one that we assume for the model parameters in Eqs. (1) and (2). The steady state solution of Eqs. (1) and (2) gives the populations of two types of states under steady illumination. In the limit where the recombination in $\gamma_{R1}$ channel is dominant (e.g. $\gamma_{R2} \approx 0$), the concentration of carriers in extended states, $n_1 \approx \sqrt{\Phi/\gamma_{R1}}$, shows effectively no temperature dependence, and cannot account for the observed photoconductivity. We therefore propose that the recombination takes place dominantly through type-2 states, which are lower in energy and generally much more populated than type-1 states. For $\gamma_{R1} = 0$ the steady state solution for $n_1$ is

$$n_1 = \frac{\Phi}{c\gamma_{12}} + e^{-\frac{\Delta}{T}} \sqrt{\frac{\Phi}{c\gamma_{R2}}}. \tag{4}$$

If both recombination channels are allowed to act simultaneously the solution is given, to a very good approximation, by the harmonic mean of the two previous solutions,

$$n_1 = \left[\left(\frac{\Phi}{c\gamma_{12}} + e^{-\frac{\Delta}{T}}\sqrt{\frac{\Phi}{c\gamma_{R2}}}\right)^{-1} + \left(\frac{\Phi}{\gamma_{R1}}\right)^{-1/2}\right]^{-1}. \tag{5}$$

We propose that the resistivity ρ(T) reflects the evolution of $n_1(T)$ with temperature through the simple Drude formula[18]

$$\rho = \left(\frac{n_1 e^2 \tau_1}{m}\right)^{-1}, \tag{6}$$

where $m$ is the effective band mass of type-1 states. We assume that the carrier scattering rate $\tau_1^{-1}$ follows the usual monotonic increase with temperature (e.g. $\tau_1^{-1} \approx a + bT$)[18], arising essentially from disorder scattering and crystal lattice fluctuations (phonons). The resulting ρ($T$) features the metallic increase at low temperatures and the exponential decrease at high temperatures. This corresponds to $n_1$ being constant at low temperatures, and exponentially increasing at higher temperatures. The former value is determined by the "inter-state" transfer rate $\gamma_{12}$, while the latter is the result of the "quasi-equilibrium" between type-1 and type-2 states that builds up at higher temperatures. The sizable transfer-rate $\gamma_{21}$ ($\gamma_{12}$) is essential for building the latter state, but $n_1$ in that state is determined only by the recombination rate and the light intensity. The linear dependence of photo-induced carriers on light intensity in the low temperature state evolves to the square-root dependence in the high temperature state, fully confirmed by the experiments (inset to Fig. 2).

It is important to realize that the experiment dictates the hierarchy of rate constants: (i) The strong rise of resistivity in the low temperature region implies that the "trapping" rate $\gamma_{12}$ is significantly lower than the scattering rate $\tau_1^{-1}$. (ii) The "trapping" rate $\gamma_{12}$, has to be second fastest one in the hierarchy. This is essential for the occurrence of quasi-equilibrium in the high temperature state and the exponential temperature dependence of the resistivity. (iii) The recombination from type-2 states is dominant in the system, $\gamma_{R2} n_2^2 \approx \Phi \gg \gamma_{R1} n_1^2$. The recombination in $\gamma_{R1}$ channel is irrelevant in the low temperature region, and marginal-to-relevant at high temperature (this is valid even for $\gamma_{R1} \gg \gamma_{R2}$, as long as the carrier densities differ greatly, $n_1 \ll n_2$); (iv) The system generally operates away from the saturation regime (the saturation , $n_2 \approx 1$, may obviously arise in a very clean material, and the extended model discussed in the supplement reaches that limit).

There are many ways to further elaborate the model in its mathematical and physical details [22]. The essential features to be maintained are a) the existence of long-lived, "charge reservoir" type-2 states and b) the hierarchy of scattering/trapping/recombination processes, as outlined in the minimal model. The minimal model reproduces the essential features of the experimentally observed photo-resistivity, as shown in Figs. 2, 3 and 4.

This leads us again to the important question about the origin and nature of the type-2 states. One can consider three possibilities, with the localised states coming from 1) defects or impurities; 2) from excitons, bound states of electrons and holes; 3) LUMO molecular levels. For the latter case one could consider the $CH_3NH_3$ level, but it is 2.5 eV above the CB minimum [19] so it is inaccessible for transport measurements. Excitons are certainly present in the illuminated material, although their binding energy and life-time are not yet completely settled[20]. If they would play a role in storing electrons, the triplet excitons states would be preferable for their generally much longer life times[21]. An argument against their involvement here is Fig 4, which shows a strong dependence of ρ on sample quality. These curves speak in favour of the extrinsic origin of type-2 states. The best candidates for these trap states are defects. In a recent paper, Kim, et al.[5] claimed that the ionic bonding nature of $MAPbI_3$ makes charge carriers very much defect tolerant, because the loss of $CH_3NH_3^+I^-$ species create neutral defects which do not appear within the band gap. Even Pb or I vacancies do not form deep levels like in standard semiconductors (e.g., GaAs, Si etc.[18]). According to their calculation these imperfections can cause, at most, very shallow levels below the CB which could be merged with the CB. Our findings that the dark current is very low corroborate with the statement of Kim et al. [5] that these defects are neutral. However, the type-2 states show a substantial distance (~0.1 eV) from the CB, have a certain spread in energy (see Fig. 4) and are populated by photo-electrons. Our model encodes the "metallic-to-semiconducting" crossover on the concentration $c$ of localized states, which is qualitatively followed in experiment. These results suggest that the remarkable feature of MAPbI$_3$ is not the absence of impurity/localized states, but instead the long lifetime of the carriers in those states which facilitates their re-entering into the CB by thermal excitation.

In conclusion, the results of photoconductivity experiments presented here show very directly that the processes like exciton formation and recombination in MAPbI$_3$ occur on time scales significantly longer than the ordinary transport scattering time. This means that electrons and holes created by light diffuse towards electrodes essentially undisturbed by processes that would reduce their concentration. Moreover, the pronounced rise of the resistivity with temperature in the low-temperature region shows that the scattering of electrons on lattice imperfections, in our simply prepared crystal, is much smaller than unavoidable scattering on thermal lattice fluctuations. The temperature dependence of scattering in the metallic region is characterized by the ratio b/a of the order of 0.1/K. This implies that already around the temperature of structural transition and even more above that temperature the scattering on crystal imperfections is practically insignificant. This, together with the hierarchy of process rates established above, also provides additional background for understanding the long excitons diffusion lengths found recently in the same compound.

Beyond photovoltaics, our findings may have further relevance. First, it should be noted that the essential processes and conditions required for the understanding our experiments (the transfer of electron from the excited state prepared by the light pump to another state that is long-lived and lower in energy) coincide with generic processes and conditions for the preparation of the inverted population state that precedes lasing, experimentally reported recently by Xing et al. [23]. Second, it may be recalled that the interest for this organic-inorganic perovskite was initiated by D. Mitzi in the early nineties [24], in search for new perovskite materials which are metallic, and eventually superconducting at high temperatures like the cuprates [25]. Current developements progresively show that the hybrid perovskites may indeed open a new era of accesible materials where collective states of electrons are created by highly tunable, clean and disorder-free doping through illumiantion.

Author information

*Corresponding Author: email: laszlo.forro@epfl.ch

Author Contributions

The manuscript was elaborated through contributions of all authors. All authors have given approval to the final version of the manuscript.

Notes

The authors declare no competing financial interest.

**Methods summary**

$CH_3NH_3PbI_3$ crystals were prepared by precipitation from a concentrated aqueous solution of hydro-iodic acid containing lead (II) acetate and a respective amount of $CH_3NH_3^+$ solution. A constant 55-42 °C temperature gradient was applied to induce the saturation of the solute at the low temperature part of the solution. After 24 hours sub-millimeter sized flake-like nuclei were floating on the surface of the solution. Large $MAPbI_3$ crystals with 3x5 mm silver-grey mirror-like facets were grown after 7 days. The structure and the phase purity were verified by x-ray and Raman scattering measurements. Leaving the crystals in open air resulted in a silver-grey to green-yellow colour change. In order to prevent this unwanted reaction with moisture the crystals were immediately transferred and kept in a desiccator prior the measurements.

For electrical contacts gold stripes were evaporated on the sample surfaces and 50 µm gold wires were glued onto using silver paste (see Fig. 1). The resistances were measured in two and four point configurations. The former one gives a factor of two higher resistance but essentially the same dependence on temperature or illumination. R was calculated using an effective thickness of 1 µm, the depth in which light is completely absorbed and photo-electrons are generated. The peak in the resistivity at the structural phase transition shows up in the 152 - 162 K, the lowest being for the sample with the highest defect density concentration c.

**Supplementary information**

1. Model extensions and the regime of photo-induced dopants:

The model exposed in the main text may be significantly extended, with some new physical regimes included. Here we comment on two extensions that may have physical relevance in the frameworks of our experiments. The rate equations in the extended model acquire additional terms to become,

$$\frac{dn_1}{dt} = \Phi_1 - \gamma_{R1} n_1^2 - c \gamma_{12} n_1 (1 - n_2) + c \gamma_{12} e^{-\Delta/T} n_2 , \qquad (E1)$$

$$\frac{dn_2}{dt} = (1 - n_2)\Phi_2 - \gamma_{R2} n_2^2 - \gamma_{12} n_1 (1 - n_2) - \gamma_{12} e^{-\frac{\Delta}{T}} n_2 . \qquad (E2)$$

First, the direct excitation of carrier into a type-2 state is introduced through $\Phi_2$, distinct from direct excitation into the extended states, already present in the main text and here denoted by $\Phi_1$. Second, the possibility for *saturation* of the population of type-2 states, $n_2 \to 1$, is introduced through the $(1 - n_2)$ factor: the transfer of carriers from type-1 to type-2 states slows down as $n_2$ approaches the maximal value of 1. Direct light-induced generation of carriers into type-2 states is also slowed down. Saturation is likely to occur in the limit of extremely low concentration $c$ of type-2 states. Especially interesting is the limit of $\Phi_2 \gg \Phi_1, \gamma_{R2}$, which leads to the regime of "photo-induced dopants". In that regime the transfer-rate from type-2 to type-1 states is greatly suppressed at low temperatures, and the saturation, $n_2 \approx 1$, is caused by the $\Phi_2$ source. The saturation effectively terminates the link between type-1 and type-2 states; $n_1$ is determined by $\gamma_{R1}$ recombination, $n_1 \approx \sqrt{\Phi_1/\gamma_{R1}}$ and independent of temperature. The transfer $2 \rightarrow 1$ gets activated at higher temperature, leading to "doping" of type-1 states from saturated type-2 states. The overall dependence of $n_1(T)$ is qualitatively similar to that discussed in the main text, but there are also some important differences. First, in the "photo-induced-dopant" regime the value of $n_1$ and the photoconductivity turn sample independent (insensitive to $c$) in the low temperature region. Second, the photoconductivity in the "photo-induced dopants" regime is characterized by very weak dependence on light intensity in the high-temperature region; the square-root dependence of the photocurrent is expected in the low-temperature regime. None of these is found in our experiments.

Various regimes that appear in the extended model were explored in great detail numerically and analytically. The model proposed in the main text is the result of these explorations, reduced to terms that appear to be the most relevant for the material.

2. Parameters of the fit of fig 3a with the expression:

$$A * \frac{a + b * T}{\Phi/c\gamma_{12} + \sqrt{\Phi/c\gamma_{R2}} * \exp\left(\frac{-\Delta}{T}\right)}$$

| $\Phi$ (lux) | a (a.u) | a/b (K) | $\Delta$ (meV) | $c\gamma_{12}$ (a.u) | $(c\gamma_{R2})^{1/2}$ (a.u) |
|---|---|---|---|---|---|
| 16000 | 42 | 11.35 | 163 | 16000 | 750 |
| 34000 | 42 | 11.35 | 143 | 20606 | 1450 |
| 44000 | 34 | 11.33 | 120 | 16000 | 2100 |

Comments on the fit parameters: 1. the a/b ratio confirms the predominance of the temperature dependent part over the residual resistivity; 2. the decrease of $\Delta$ with $\Phi$ indicates that type-2 states that are lowest in energy tend to progressively saturate as the light intensity increases; 3. the monotonic variation of $(c\gamma_{R2})^{1/2}$ suggest that with higher light intensity the recombination through $\gamma_{R1}$ (omitted in the basic analysis) increasingly plays its role.